# Robust Identification of Email Tracking: A Machine Learning Approach

**Johannes Haupt**[a*], **Benedict Bender**[b], **Benjamin Fabian**[c], **Stefan Lessmann**[a]

[a] School of Business and Economics, Humboldt University of Berlin,
  Spandauer Str. 1, 10178 Berlin
[b] Chair of Business Informatics, University of Potsdam,
  August-Bebel-Str. 89, 14482 Potsdam
[c] Chair of Business Intelligence and Data Science, Hochschule für Telekommunikation Leipzig,
  Gustav-Freytag-Str. 43-45, 04277 Leipzig

## Abstract

*Email tracking allows email senders to collect fine-grained behavior and location data on email recipients, who are uniquely identifiable via their email address. Such tracking invades user privacy in that email tracking techniques gather data without user consent or awareness. Striving to increase privacy in email communication, this paper develops a detection engine to be the core of a selective tracking blocking mechanism in the form of three contributions. First, a large collection of email newsletters is analyzed to show the wide usage of tracking over different countries, industries and time. Second, we propose a set of features geared towards the identification of tracking images under real-world conditions. Novel features are devised to be computationally feasible and efficient, generalizable and resilient towards changes in tracking infrastructure. Third, we test the predictive power of these features in a benchmarking experiment using a selection of state-of-the-art classifiers to clarify the effectiveness of model-based tracking identification. We evaluate the expected accuracy of the approach on out-of-sample data, over increasing periods of time, and when faced with unknown senders.*

**Keywords:**  Analytics, Data Privacy, Email Tracking, Machine Learning

---

[*]Corresponding author: johannes.haupt@hu-berlin.de

# 1. Introduction

Data on email reading behavior is routinely used to infer commercially valuable information from customers. For example, it allows marketers to derive user profiles and measure the reach and effectiveness of email marketing campaigns (Hasouneh & Alqeed, 2010). It also facilitates marketing activities, such as calling prospective customers at the time they open a marketing message (Hlatky, 2013). The Direct Marketing Association estimates that its members achieved an average return of £38 for every pound spent on email marketing and that this ROI will continue to increase in the future with the spread of advanced testing and personalization (The Direct Marketing Association, 2015). This gives marketers a strong incentive to monitor how customers interact with email newsletters and advertising. Using the same methods, spammers and phishers rely on email tracking to validate and collect active email addresses for their illegal activities (Sophos, 2014). Current email tracking techniques enable the sender to track if and how often an email is opened, the time at which the email is read, which device as well as operating system the recipient uses, and her Internet Protocol (IP) address (Murphy, 2014). Such information, in turn, facilitates deducting the location of the reader, her affiliation to a company or organization, email reading behavior, travel patterns based on desktop and mobile use, and if an email was forwarded or printed (Technology Analysis Branch, 2013). A peculiarity of email tracking is that tracking information is linked to a user's email address, which is an almost unique identifier of the user that can easily be matched to other accounts of the user such as social media profiles. Consequently, tracking users across devices, applications, locations, etc. is much easier in email tracking compared to other channels such as web tracking. Importantly this data is typically gathered without active consent, case-by-case confirmation or even awareness of the recipient. In combination, these characteristics facilitate surveillance and constitute an invasion of user privacy. As we are able to show, email tracking does not merely constitute a theoretical risk but is ubiquitous in marketing communication.

Therefore, email users require tools to protect against potential privacy hazards caused by email tracking. A review of the literature and contemporary email clients reveals a lack of easy-to-use, effective, and reliable protection methods. The reason is that the identification of tracking images, which are the main tracking mechanisms in emails, poses specific challenges that render standard ad blockers and blacklists ineffective. The goal of this paper is to contribute towards empowering email users to protect their privacy. To that end, we develop a machine learning approach to detect tracking elements in emails with the ultimate goal to filter them selectively.

The contribution of this paper is three-fold. First, we establish the prevalence of email tracking through the analysis of 30,756 marketing-communication emails from 300 global companies collected over a period of 20 months. We extend previous analyses by comparing the occurrence of email tracking in different industries and identifying common email-tracking providers. Second, we develop a set of features geared towards the identification of tracking images under real-world conditions. These features are devised to be computationally efficient, to generalize to structures of unseen tracking images, and to be resilient against changes in tracking structures over time. Third, using a selection of state-of-the-art classifiers, we test the predictive power of these features in a benchmarking experiment to clarify the effectiveness of model-based tracking identification. We evaluate the expected accuracy of the approach on test sets that are out-of-sample, out-of-time, i.e. after increasing amounts of time have passed, and out-of-universe, i.e. when faced with unknown senders. This allows us to identify an optimal identification model and appraise the degree to which a model-based approach protects against email tracking in application.

The remainder of the paper is structured as follows. Section 2 introduces current email tracking techniques. Section 3 identifies related literature. Section 4 examines the occurrence of tracking within the commercial newsletters that we collect for the study to stress the relevancy of defensive strategies. Section 5 presents the featurization methodology to identify tracking images. Section 6 and Section 7 elaborate on the experimental design and empirical results, respectively. Section 8 concludes.

## 2. E-Mail Tracking Technology

We start by outlining email tracking methodology and the degree to which it impacts user privacy. This section provides the technical foundation to develop features for tracking identification and countermeasure design. The tracking process (Figure 1) is based on emails that are written in Hypertext Markup Language (HTML) referencing specific external resources. Prior literature refers to these resources with different terms, including "web bugs" (Martin, Wu, & Alsaid, 2003) and "tracking pixels" (Vaynblat, Makagon, & Tsemekhman, 2009). Considering their function and location within the email's HTML code as  tags, we use the term *tracking images*. The tracking process starts with the sender dispatching an HTML-based email. The email includes an image tag, which references a tracking object stored on a server of the sender, or its tracking provider, in the form of a Uniform Resource Locator (URL). When the recipient opens her mail client, the mail user agent (MUA) synchronizes the local mail repository with updates provided by the recipient's message transfer agent (MTA), and the user receives the email. When the recipient opens the email, which contains the tracking image tag, the mail client requests the referenced file. The web server, where the file is stored, logs this request and provides the image to the client. Log analysis allows the sender to infer information on the recipient's access device and email reading behavior. For example, if the email is opened on different devices, every individual access is logged with the corresponding user-agent information, which allows for cross-device tracking.

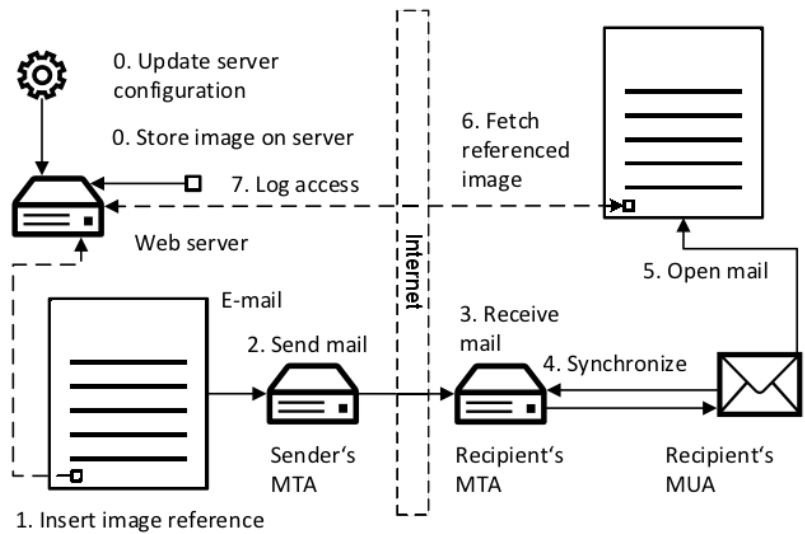

**Figure 1: Overview of the email tracking system and process**

Image requests themselves do not contain sufficient information to identify a specific email recipient. For the purpose of matching an image request to a known recipient and thus track individual behavior, either the tracking object must be unique to the recipient or the reference URL must contain a unique tag that identifies the recipient. In both cases, the reference within the  tag will be unique to the email recipient. By sending images with a specific reference to only one recipient, trackers control that subsequent access to the image via that reference can be attributed to a single recipient. The hash of the recipient's email address has been identified as a common approach to create anonymized identifiers (Englehardt, Han, & Narayanan, 2018).

In contrast, requests to non-tracking images from references that are common to all recipients, e.g. product pictures, are logged on the server in combination with the respective IP address and device information but cannot be linked to recipients' email addresses. An extension to this form of aggregate data collection are images containing, for example, an identifier for the email campaign rather than individual recipients. Use cases of tracking on the aggregated level (i.e., without an identifier for individual users) include measuring the opening rate of an email campaign for A/B testing of newsletter design. Since no individual information is collected by non-unique tracking images, their privacy implications are less pronounced. We consequently focus on images that include a unique identifier and facilitate tracking of individual users in this study. For readability, we refer to individual tracking images as tracking images.

Individual tracking data poses a privacy risk because personal information about the identity and behavior of the tracked user can be derived without her consent or awareness. The log entries facilitate inducing that the user has read or at least opened an email, because current email clients do not download images before the corresponding email is opened. In case of spam emails sent to random email addresses, this is sufficient to prove that an active account has been found. In addition, the time stamp and the existence of multiple log entries reveal the time of day and the number of times an email is opened. The combination of multiple entries for a single mail, as well as multiple entries from one user for different mails, provide insights into the recipient's email reading behavior. Furthermore, the log entry facilitates inferring information about the user environment (Agosti & Di Nunzio, 2007). Data on the use of mobile or desktop devices, especially when aggregated over time, conveys additional information about user activities such as office or travel times. It is also possible to track whether an email has been printed through a print stylesheet, either by tracking the stylesheet directly or by matching stylesheet access to the device information collected by the tracking image.

More complex analysis reveals additional information about the recipient. For example, transmitted IP addresses enable trackers to gather location-related information (Poese, Uhlig, Kaafar, Donnet, & Gueye, 2011). Based on a reverse lookup of an IP address, a log entry may also reveal a user's affiliation to an organization, for example, if private emails are opened at work. Combining pieces of information also facilitates predicting whether an email has been forwarded and allows deducting travel routines. For example, a major technology company combined the IP address, location information and the time stamp of a log entry to identify a board member who was forwarding confidential information (Evers, 2006).

A crucial point differentiating web and email tracking is that the collected and combined information is not anonymous in email tracking. While both rely on similar mechanisms (e.g., cookies or tracking images) and gathers a rich set of behavioral information, users tracked via web tracking are not directly personally identifiable without consent. The personal identification of the tracked user is often impossible and alternative methods to recognize users over time and web sites have been proposed (Nikiforakis et al., 2013) (Yang, 2010). Information collected via email tracking, on the other hand, is necessarily linked to an email address, which provides a platform independent almost unique identifier of a person and often contains the user's name and possibly organization. Additionally, it is often possible to link an email address to personal online profiles, for example on social media sites.

Currently, the only solution for providing fully reliable privacy protection against email tracking in HTML emails is to block all external content referenced in emails. From a technical point of view, this approach is easy to implement on either the server or the client and can be activated as default for most email clients. However, blocking all images in an email entails a substantial loss of information and interferes with user experience by excluding all referenced images and the corresponding content. Possible further issues include incorrect formatting, loss of styling elements, and misinterpretation if external images convey crucial information.

A selective filtering approach provides a balance between preventing user tracking and sustaining user experience. It operates through identifying and selectively blocking tracking elements within an email. In this approach, a predictive model is used to categorize referenced images into tracking and non-tracking images. Non-tracking images remain untouched, whereas tracking image references are removed from the email. Note that tracking images are often transparent and do not contain content (Bender, Fabian, Lessmann, & Haupt, 2016). In the ideal case, the user avoids being tracking without noticing that an email has been sanitized. However, the efficacy of selective filtering depends critically on the algorithm for tracking image identification.

## 3. Related Work

We organize the literature related to this study into three categories. First, we summarize the existing research on email tracking. Given the sparsity of research on this specific topic, we next identify studies on web tracking, which is similar from a technological perspective. Last, we discuss previous studies investigating mechanisms to selectively remove unwanted elements from HTML-based content.

Email tracking is periodically covered in the general press, where it is criticized for invading privacy (Murphy, 2014) or mentioned as a tool to uncover information leakage (Hodgekiss, 2010). Some authors hint at the possibility of tracking in HTML emails (Bouguettaya & Eltoweissy, 2003; Harding, Reed, & Gray, 2001; Martin et al., 2003; Moscato, Altschuller, & Moscato, 2013; Moscato & Moscato, 2009). Few academic papers have examined the topic. A notable exception is the recent study by Englehardt et al. (2018) showing the ubiquity of email tracking in a large scale sample. Most studies focus on marketing rather than privacy or countermeasures against email tracking, for example Bonfrer and Drèze (2009) and Hasouneh and Alqeed (2010), who structure technical and process-related aspects of email tracking from a marketing perspective and stress the importance and prevalence of tracking in newsletters and other marketing communication. This study extends our own previous research on the characteristics of email tracking images as well as mechanisms for tracking detection and prevention (Bender et al., 2016) in three ways. First, we broaden the scope of the analysis of tracking prevalence through examining emails gathered over a horizon of 20 months and from 33 industries. Second, we substantially improve the tracking detection engine. Whereas Bender et al. (2016) use an untuned feed-forward neural network classifier, we conduct a comprehensive benchmark of state-of-the-art machine learning algorithms for tracking image classification. Third, we propose novel predictors of email tracking to ensure deployability. Most importantly, we establish the stability of detection accuracy and generality of the tracking protection framework through rigorous out-of-time and out-of-universe testing. This allows us to demonstrate that the proposed system is adequate to protect users against privacy invasions under real-world conditions.

From a technological point of view, email tracking can be considered an adaptation of web tracking mechanisms to HTML-based emails. Unlike email tracking, the use of web tracking in different situations (Javed, 2013; Jensen, Sarkar, Jensen, & Potts, 2007) and its detection (Alsaid & Martin, 2002; Fonseca, Pinto, & Meira, 2005) have received much attention in the literature. Prevention of such mechanisms and the evaluation of existing software solutions have also been studied (Fonseca et al., 2005; Leon et al., 2012). Other research emphasizes the technical aspects of web tracking, such as different categories of web bugs (Dobias, 2010) or the potential for aggregating multiple server log files (Evans & Furnell, 2003). We make use of the mature research towards the detection of web tracking and extend it to email tracking.

Methodologically, the identification of tracking content is related to the identification of tracking and advertising on web pages (e.g., ad blocking) or other unwanted content in emails (e.g., spam and phishing detection). These applications make use of information related to the image reference URL, the email sender, the website host, the content visible to the user, and the formatting of an image. Ad blockers rely on the image content for classification (Li, Li, Li, & Wang, 2011). However, content classification requires accessing the image, which would be registered by the tracking server. Therefore, content-based approaches are inapplicable to prevent email tracking effectively.

An alternative is to examine the structure of image references. Li et al. (2011) and Kushmerick (1999) propose a range of features to identify advertising images on web pages. They focus on the formatting and image-reference link relative to other images on the same page; for example, investigating whether the image domain is different from the site domain, with a deviation being indicative of third-party content. The reference structure itself is also used to identify advertisement. For example, Shih and Karger (2004) propose a heuristic that exploits the fact that advertisement images are often placed in a different folder than content images. URLs have also been used with success to identify phishing mails (Blum, Wardman, Solorio, & Warner, 2010; Garera, Provos, Chew, & Rubin, 2007; Ma, Saul, Savage, & Voelker, 2009b; Whittaker, Ryner, & Nazif, 2010) and to classify web pages (Kan & Thi, 2005; Shih & Karger, 2004).

Most of the above approaches rely on identifying keywords through text mining on parts of the URL. These keywords include both words in natural language describing the target-link content, meaningful letter or number combinations called tokens, and recurring server or folder names. While these can be identified for tracking images, they require constant updating and are susceptible to avoidance strategies by spammers and trackers, respectively. Fette, Sadeh, and Tomasic (2007) introduce predictors counting the number of dots and the number of different top-level domains in mail links to capture the complexity of the URL and the increasing number of domains involved in phishing. We extend these ideas when creating features to capture the structure of tracking image references.

Especially for phishing analysis, some approaches rely on the content of the email. Bergholz, Chang, Paass, Reichartz, and Strobel (2008) propose features based on a dynamic Markov chain and topic models based on Latent Dirichlet Allocation. We focus on tracking image identification but acknowledge that a pre-classification of emails based on their subject line or content could convey some preliminary information on the probability of an email being tracked. Preliminary classification could increase speed and accuracy of tracking identification in future work.

Host information has been found effective in phishing and spam detection (Fette et al., 2007; Ma, Saul, Savage, & Voelker, 2009a; Ma et al., 2009b). This information is gathered via the IP address and a WHOIS request to the domain of the server that hosts a referenced website, because a phishing site "may be hosted in less reputable hosting centers, on machines that are not conventional web hosts, or through disreputable registrars" (Ma et al., 2009b). This reasoning does not hold for email tracking in e-commerce, where businesses operate within legal bounds and tracking images are hosted on official company or contractor servers. Moreover, looking up external information slows down the identification process in potential real-time applications (Blum et al., 2010).

In summary, prior work in the context of web tracking mentions the existence of email tracking, hints at tracking methods, and criticizes privacy implications. However, we find a lack of research investigating the prevalence of (legal) tracking activities and approaches to email prevent tracking. While there exist initiatives to develop anti-tracking software in the form of modified mail clients and add-ons that support selective tracking prevention (Barret, 2015), these tools are unable to provide reliable protection against most tracking approaches (Bender et al., 2016). Therefore, we extend prior work through studying a more comprehensive set of data and providing the foundation of a detection system to identify and selectively block tracking images. Specifically, we build on existing predictors of tracking use and extend these so as to ensure feasibility and improve resilience in real-world applications. Our empirical analysis then establishes the best learning algorithm for the task of tracking image detection and estimates its performance on emails from senders not seen in the data, and also after periods of time between training and application.

## 4. Data and E-Mail Tracking Usage

Analyzing the occurrence of tracking and training a supervised learner for automated detection require data on email communication including the status (tracking/non-tracking) of every image across all emails in the data. The collection of this ground truth data is complex due to important differences in data collection between ad blocking or spam detection and the identification of tracking images. In particular, comparable studies obtain status labels through human judgment, which is often crowd-sourced. The information available for classification in our setting consists of the images themselves and the email source code. Identification of tracking images based on the image content is unreliable, since content and tracking functionality are independent of one another. In practice, transparent or tiny images without actual content are also legitimately used for formatting purposes (Martin et al., 2003). Thus, ground truth classification must be based on the image tag in the email code and, most importantly, the image reference. Image references do not have to be human-understandable, and tracking images are hidden from the recipient by design, which makes identification through human judges unreliable; as illustrated in Table 1.

**Table 1: Example image tags of two tracking and non-tracking images, respectively. Tracking image tags are shown in rows 2 and 3.**

| |
|---|
| (1)  |
| (2)  |
| (3)  |
| (4)  |

A constituent property of personal tracking image references is that they contain a unique identifier for the recipient (see Section 2). We therefore create two identities and corresponding email addresses using Gmail and match the emails and images received on both accounts to identify tracking elements. We do this by extracting images from the HTML content of each pair of emails sent to both accounts and comparing the image reference URLs at each position for differences. Images for which the reference URLs are an exact match are classified as non-tracking images and images with different URLs as tracking images. To avoid bias from senders changing their email policy in response to the reading behavior of the users they are tracking, we ensure that none of the external images are requested from the web server at any point.

With each account, we signed up for the newsletters of 300 companies and collected emails in a 20-month period from 2015 to 2017. Although not representative of email communication in general, we argue that newsletter emails are a suitable vehicle for this analysis. First, it is likely that companies use email tracking to assess the effectiveness of their newsletters (Hasouneh & Alqeed, 2010). We aim to increase this likelihood by concentrating on large companies, which are on average faster to adopt novel technology (Premkumar & Roberts, 1999). We sign up to email newsletters from the top-100 companies ranked by revenue in Germany, Great Britain, and the United States. Second, the wide availability of different newsletters simplifies systematic data collection and facilitates the gathering of a large amount of data. At the same time, signing up to newsletters requires an active request restricting the amount of data and its variance and may introduce selection bias, as opposed to, for example, the passive collection of unsolicited emails in spam detection. To mitigate this effect and ensure substantial variance, our company selection is based on company size and includes companies from three countries. Third, newsletter can be ordered multiple times without difficulty. In contrast to for example personal communication, using commercial newsletters is an effective way to gather ground-truth data.

Each artificial identity received 30,756 emails, which we could match between accounts. Of these, 7,154 (23%) are in plain-text format, while the remaining 23,602 are HTML-based and thus facilitate tracking. Of the HTML emails, 21,500 (91%) contain a total of 794,519 external image references, which constitute the data set on which we build and test the tracking detection model. The number of images per email varies considerably and shows positive skewness. We observe a mean (median) value of 37 (18) external images per email. 16,410 emails (69% of HTML emails) contain tracking elements, which illustrates that tracking is common in company newsletters. The ratio of emails received from each country roughly corresponds to the ratio of companies with 29% of emails sent by companies from Germany, 40% from the United Kingdom (UK), and 31% from the United States (US). The tracking quota and the fraction of HTML-based emails vary significantly between countries (see Figure 2). The ratio of HTML emails is close to 100% for Germany and the US. In the UK, only 44% of emails are in HTML format, and out of these, only 46% are tracking mails, resulting in an overall tracking quota of 20%. This is significantly lower than the tracking quotas in Germany (95%) or the US (69%).

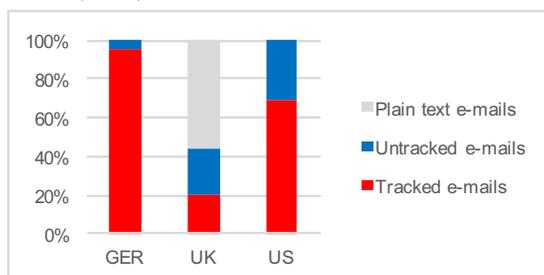

**Figure 2: Ratio of tracked emails per country**

Country-level variation reflects the industry distribution of the top companies in each of the three countries. Each email is matched to a company according to its sender domain and assigned to an industry category based on the Financial Times Equities database (Financial Times, 2017). Figure 3 presents the per-industry tracking ratio for industries with more than 100 emails in the sample. We observe that customer-targeted newsletters are tracked with near certainty, while business-to-business newsletters and company news, predominant among industrial producers, are less likely to contain a tracking image. An exception to this rule are investor bulletins, which are sent at high frequency in plain text. Bulletins are responsible for the large fraction of plain-text emails (light grey color) observed for the banking and travel sector.

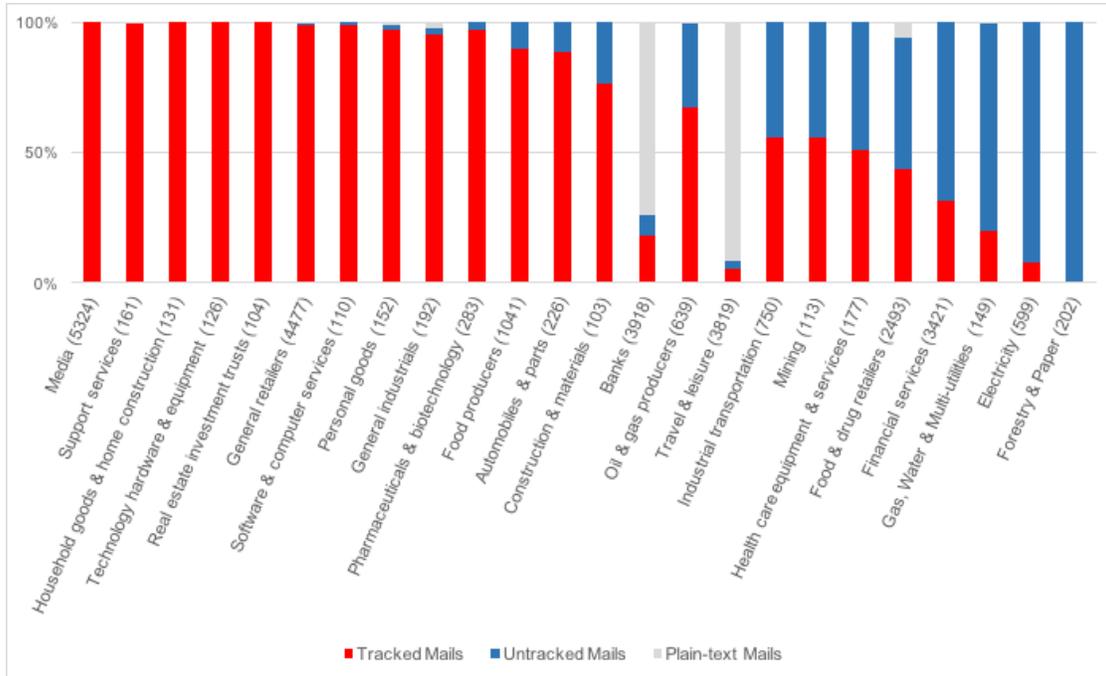

**Figure 3: Ratio of tracking by industry (total number of emails in brackets) showing industries**

The tracking literature assumes tracking images to be small, typically with an area of 1 square pixel (Martin et al., 2003). We analyze the observed image sizes for tracking and content images in Figure 4. 35% of the tracking images for which a size could be determined have an area of one square pixel. There exist images with a specified area of 0, which are most likely not shown by the email client thus making them effectively invisible. The majority of tracking images has an area above 100 square pixels (38%) or no specified size (13%). Note that we consider several ways to specify the size of images (see Section 5.1). The results suggest that simple rules to filter images based on their area are likely to fail.

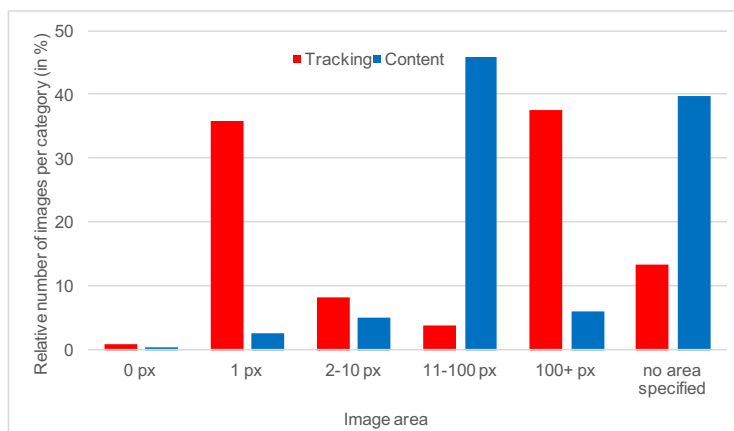

**Figure 4: Image area (height x width) for tracking and content images**

The file extensions extracted from each image reference (Figure 5) reveal that the file format is not indicated for two-thirds of tracking files. Approximately 20% of references including a file format indicate the file to

*10*

be a code script rather than an actual image file, with the majority of scripts written in PHP or ColdFusion Markup. The use of executable files instead of images sheds light on the underlying tracking infrastructure and suggests that the file access and the information associated with it can be dynamically processed or forwarded to internal or third-party databases. The findings suggest the file type – when available – is highly discriminatory for the identification of tracking images.

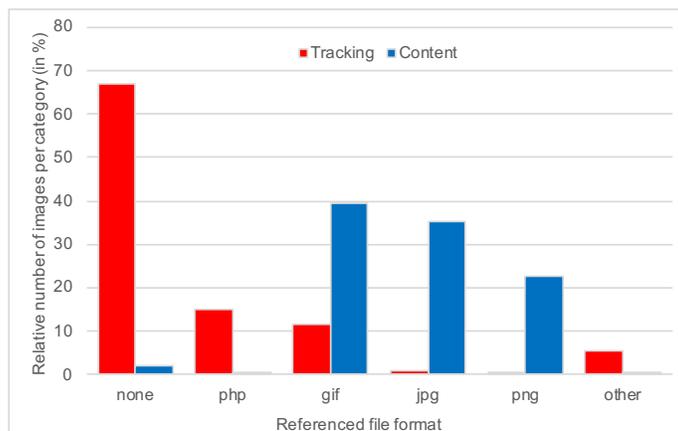

**Figure 5: Relative frequency of file formats for tracking and non-tracking images**

## 5. Tracking Image Detection

A selective tracking prevention system that targets and filters tracking elements conceptually consists of three components. First, a data-input component extracts all image tags and their attributes from raw email code. Second, a tracking detection engine, which represents the core of the system, performs two tasks. It creates indicative features from the raw data (e.g., HTML image tag) and uses the features as input to a classification model. The model estimates the probability that an image is used for tracking. Third, the selective filtering component processes the estimated tracking probability to handle external images. Rather than blocking all images in an email, which is currently the most secure way to avoid email tracking, the system is able to selectively block the download of images with high tracking probability. This empowers users to see uncritical content without being tracked and to decide, after inspecting a sanitized version of an email, whether they want to permit the download of further, system-filtered images, despite the risk of being tracked. This way, the envisioned system also offers a viable approach to handle content images that perform tracking. More specifically, users are enabled to make a conscious decision how they trade-off the risk of being tracked by the sender of an email with possible readability issues caused by image filtering.

The detection engine classifies unknown images into two categories, *tracking* and *non-tracking/content* on the basis of meta-data extracted from HTML code. Images used for individual tracking exhibit structural peculiarities, which facilitate such classification. In particular, they contain a unique user identifier assigned by the tracker, are distinctly formatted, and are often handled by a different department or company (Bender et al., 2016). However, correct identification is challenging – even for human judges – for two reasons. First, tracking images do not necessarily fulfill all criteria simultaneously and show significant variation in the observed patterns. While certain structures are necessary or common, their actual format depends largely on choices made by the tracker. For example, the existence of an individual identifier is necessary, but the identifier itself may be constructed out of numbers, lowercase and uppercase letters or any combination thereof and its position in the reference can be as folder name, image name or URL parameter (see Table 1). Note that the  attribute itself is not transmitted within the request to the webserver and thus not suitable for user tracking. The identification of tracking images is further complicated by the possibility to track images of any format or size, including branding or content images. Second, non-tracking images may display the above characteristics, including very small images used for formatting, or images handled through content-management systems, whose file names resemble user IDs. Under these restrictions, only complex rules can ensure satisfactory identification of unwanted images at a low rate of false identifications without interfering with the email content or formatting. We therefore employ machine-learning techniques to develop a detection model.

In the remainder of the section, we propose a set of features to serve as input to a supervised machine-learning algorithm. Recall that we perform our analysis at the level of an individual image. The features are split into four categories (see Table 2). The first two categories, *reference structure* and *HTML image attributes*, subsume aspects that are directly associated with the formatting of the image within the email and its reference URL path. The category *image server* is associated with the servers that host the images. The fourth category covers the *email header*. Features found in prior work (Bender et al., 2016) are marked with an asterisk. While further extension of features is surely possible, we aim to show that a set of resilient features is sufficient to ensure a high level of privacy. We elaborate on the empirical performance of the features in Section 7.1.

**Table 2: Predictors for the detection of tracking images by category**

| Reference structure | Reference structure (cont.) | Email header |
|---|---|---|
| Count IDs in filename * | Reference includes '?' | *Custom header fields* |
| Count IDs in path * | Reference includes '@' | Image name matches sender |
| Count number strings * | Reference includes 'id' | Length 'unsubscribe' field * |
| Count number-letter changes * | Reference includes 'click' * | Ref. parts match 'list-unsubscribe' * |
| Count numbers * | Reference includes 'open' * | Ref. parts match 'received-spf' * |
| Count punctuation * | Reference includes 'track' * | |
| Count strings | Reference includes 'view' * | **Image server** |
| Count uppercase * | | Images sharing same domain |
| Fileformat 'jp(e)g' * | **Image structure** | Matching image and sender domain * |
| Fileformat 'php' * | *align* * | |

| | |
|---|---|
| Fileformat none * | *Area* * |
| Img. sharing same fileformat | *Area: 0 pixels$^2$* * |
| Filename length | *Area: 1 pixels$^2$* * |
| Image link similarity (Max.) | *Area: 100+ pixels$^2$* * |
| Image link similarity (Mean) | *Area: 11-100 pixels$^2$* * |
| Image link similarity (Min.) | *Area: None specified* * |
| Image reference (ref.) length * | *Border width* * |
| Rel. reference length | *Length of attribute 'class'* * |
| Length of domain | *Contains 'style: display'* * |
| Longest Number in reference | Count other identical images * |
| Difference to mean letter count 'b' | *Image width* * |
| Difference to mean letter count 'f' | *Length of tag 'Title'* * |
| Difference to mean letter count 'm' | Ratio of smaller images |
| Difference to mean letter count 'w' | Rel. image position * |
| Number of folders in path | |
| Rel. filename length | |
| Rel. number of folders in path | |

*Features in italics are excluded from model training to ensure generality and resilience*

\* Features marked with asterisks have been introduced by Bender et al. (2016)

In the following, we refer to the task of creating features for a detection model as featurization. Featurization is guided by the analysis of the differences in non-tracking and tracking images and domain knowledge regarding the tracking process. We extend the features from Bender et al. (2016) and select a subset of features for model building based on theoretical considerations of generality and resilience, where resilience describes features and models with stable performance in the event of potential defensive strategies by trackers and changes in tracking infrastructure. It is reasonable to anticipate that companies and tracking providers will adjust their tracking infrastructure to evade anti-tracking efforts; similar to the efforts of spam senders to outsmart spam filters. We expect generality to require features capturing common and inclusive patterns and resilience to require features that cannot be effectively modified by trackers. Two general strategies are applicable to achieve this goal. Based on our understanding of the tracking process, we first exploit the user identifier as an observable and necessary trace of the tracking method and develop features that comprehensively describe its common form as a hash or random letter-number string. The goal is to determine a range of characteristics that are sufficiently general to be prohibitively costly or technically impossible for trackers to avoid. Second, we relate characteristics of single images, which we derive from the data or the related studies on web tracking and ad detection, to other images within the same email. By evaluating each image within the context of the email, potential adjustment strategies by trackers need to consider the infrastructure and conventions used by the content handler. While we engineer and select features based on domain-knowledge and theoretical considerations, future approaches could monitor possible patterns of misclassifications and actual tracker reactions.

## 5.1. Image structure

Image structure features are attributes that are directly associated with an image element and those referring to centrally defined style information from Cascading Style Sheets (CSS). For this category, featurization disregards HTML image attributes occurring in less than 1% of the images in our data set to avoid rare and custom tags and ensure that patterns are detectable and relevant. For tracking images, we expect image attributes to leave display options undefined or make the image harder to detect. For example, a manual inspection of a small set of tracking images suggests the attributes *border* (i.e., the thickness of the border around an image), *style properties* and their respective CSS commands, *vspace* and *hspace*, (i.e., white spaces around images) to have a good discriminatory power (Musciano & Kennedy, 2006).

We further account for the total *number of images* and *relative position* of each image within an email. Our data exploration shows tracking pixels often occur as the first or last image in the email. We suspect that tracking software automatically appends the tracking image to the top or end of an outgoing email to not disturb the email content and furthermore is easier to implement if outsourced tracking services are employed. A second aspect is related to the number of *occurrences of each unique image* within an email. Images used for formatting or branding may be used more than once in one email, but there is no technical nor functional reason to reference the tracking image in an email several times.

A very small image size is often regarded as a typical characteristic of tracking images. There are several ways image size can be specified. The  attributes *width* and *height* allow direct specification of the size of the displayed image when the website or email is rendered (Musciano & Kennedy, 2006). Height and width can also be set in the *style* option, sometimes as a maximum value or in relation to its parent block, or only one dimension can be specified, in which case the image is resized with fixed ratio. It is also possible to not set any size to display the image in its full size. Where no size is explicitly set, we try to extract the image size from the file name, where it is often indicated in the form *image_180x120.gif* or similar. Nevertheless, there remain both content and tracking images for which no size information is available, which are classified as "no area specified" (see Figure 4). However, there are several theoretical arguments to avoid classification based on image size. First, any image can be tracked independently of size and content. Since there is no technical restriction for tracking images to be of a specific size beyond saving server space, it is likely that tracker will adjust or randomize image size. Second, not all images below a size of 10 square pixels are used for personal tracking. Small or invisible images are also used for the design or formatting of the email content and false classification of these could corrupt display of the

email. We consequently exclude all image size features from the models with exception of the ratio of *smaller images within the same email*.

## 5.2. Reference Structure and Content

The majority of the features we propose relate to the referencing link that points to the image (i.e., the URL) with two goals. First, features describe the general structure of the reference to detect patterns that differ from the other image references within the same email, which suggests a third-party tracker. Second, features capture patterns that suggest the existence of a user identifier. Each tracking image reference necessarily contains a unique user ID in the image reference (see Section 2) in order to match the image access to a specific email recipient. While the identification of the particular ID of a single user is useful only within the context of the user and the specific sender, there is large potential in features that identify the characteristics of user ID and are resilient to changes by the tracker. In order to capture a range of possible ID structures, we create features that describe the characteristics of the reference path, the content in terms of the reference as a string, and the similarity of each reference to other images in the same email.

The reference structure is captured by a set of features targeting the link folder tree and the characteristics of each of its elements. In addition to the total length and number of elements, we further break up each element in the file path based on punctuation characters. This allows us to collect the characteristics of sub-domains and the referenced files. The observation that the vast majority of tracking images are different from the content images in each mail, which in turn tend to be similar to each other, motivates featurization to capture the *similarity between references in the same email*. We measure link similarity by the Ratcliff/Obershelp text distance between reference URLs (Ratcliff & Metzener, 1988). This text similarity has the property that identical ID tags between references tend to substantially have a high impact on the similarity value due to their relative length. To better capture structural similarity, we additionally quantify the deviation from the majority of references in the same email on several of the features discussed above, including *relative reference length* and *relative path depth*.

A direct approach to flag user identifiers within image references is to blacklist keywords that indicate tracking functionality. We can identify keywords through text analysis of the references by defining each reference link as a *bag of words* separated by punctuation or special characters and filtered for rare terms. We construct five binary features indicating the existence of tokens that have the highest ratio of occurrence in tracking vs. non-tracking images, such as *uid* or *open,* following the idea is that at least parts of the reference are usually human-readable for convenience. In cases where no random or hashed identifier is used, an @-sign within the URL identifies cases where the email address of the recipient is used as a user identifier directly. While predictive, any specific keywords exist for convenience only and are easily altered or omitted by trackers. Keyword features are consequently excluded from the model features.

A resilient heuristic for ID-like structures is to count the *number of specific special characters* that fulfil a technical role in the tracking infrastructure. In particular, parameters like the user and campaign identifier are passed to tracking scripts through the reference URL. The URL structure required to correctly parse the parameter is defined in public standards (Berners-Lee, Fielding, & Masinter, 1998). The parameters are included behind the file name after a question mark with each key-value pair linked by an equal sign. In contrast to arbitrary keywords, these characters are a necessary component of the tracking infrastructure.

Detection of the structure of identifiers is feasible by counting the occurrences of patterns in the *sequences of upper-/lowercase letters and numbers* and the *distribution of single letters*. These are motivated by the observation that hashes and randomly created image and file names as well as user IDs are expected to contain patterns, e.g. multiple changes in capitalization, and letters that are less common in human-chosen terms. To this end, we create features that capture the difference between how often a letter occurs within a reference to the average number of time the same letter occurs within references of the same email. While these characteristics are within the control of trackers, the design of user IDs which avoid the range of the features requires prohibitive effort. To reduce the set of variables based on letter distribution, we employ preliminary testing using a random forest model on a subset of the data and select the letters *b, f, m,* and *w* as the only predictive letters based on the variable importance score described in Section 7.1.

### 5.3. Image Server

External tracking providers regularly host tracking images on their own servers, while content images are likely hosted by the email sender. Even within the same company, we expect images to regularly be provided by different subdomains depending on the process owner. This is supported by our sample, in which more than half of the servers do not host any tracking images. About one-third of the unique domains host tracking images only, while the remaining servers were observed to host both types of images. We capture information on the servers sending the email and hosting the referenced images without restricting the features to specific servers occurring in the collected data. To achieve this, we extract the *ratio of images that share the same domain* and whether the *image host matches the email sender*.

### 5.4. Header Components

An email is composed of an email body and an email header. The email header contains technical information usually not visible to the end user as well as the sender name, address and subject line. An indicator for a *match of the sender name and the image name* aims to capture consistency between the sender and image host. Analysis of the data also shows that a single ID can be used to identify a user or specific message for tracking and to associate unsubscribe requests or email replies with a recipient. In these cases, the respective header fields and the tracking image reference contain an identical ID string. We consequently create features that indicate if parts of each image reference match the content of the header fields *List-unsubscribe, Return-Path,* and *Received-SPF*. These features exploit that one user ID may be used to identify a user in different parts of the infrastructure. While the relevant parts of the sender's infrastructure may lie within the control of the tracker, sufficient changes to the infrastructure will likely be complex and costly.

**5.5. Server Black-/Whitelisting**

Server black- and whitelisting plays a significant role in advertisement and spam detection (Cormack, 2008). In the context of email tracking, the elements of the lists are the image servers that are referenced in the emails. This is an important difference to SPAM classification, where usually the sender or mail-transfer agent is the object of investigation. Although the data suggest that to block images from servers that have hosted a high ratio of tracking images in the past could be an effective way for identifying tracking images in the data set, the identified servers do not generalize to other companies and potentially not even to one company over a longer period of time. Potential exceptions are third-party tracking services. Since these services take full control of tracking image creation and hosting, their servers show the same pattern, independently of the specific client (see Table 3).

To avoid overfitting the classification model to our specific data set, we exclude the identified blacklist and the server locations from model training. Instead, we propose that the images could be filtered according to a black-/whitelist in combination with automated detection or prior to the application of the classification model with the additional benefit of reducing the number of images that need to be classified by the model. The drawback is that these lists are specific, quickly outdated, and require high maintenance effort (Ma et al., 2009b). We use the above list as baseline in the empirical tests below with the caveat that the blacklist could be extended by a comprehensive analysis of the tracking service market in general, which is beyond the scope of this study.

Table 3: Identified tracking service providers and their tracking reference structure

| Third-party tracker | Typical reference structure |
|---|---|
| Acxiom Digital | http://open.delivery.net/o?[ID] |
| Artegic AG | http://[CLIENT].elaine-asp.de/action/view/[ID]/[...] |
| Conversant (former Dotomi) | http://ads.dotomi.com/cookieredir/[CLIENT[/[...].php?[ID]=1 |
| Doubleclick (Google) | http://ad.doubleclick.net/ad/[…]/[…];ord=[ID];u=[…]? |
| Mailchimp | http://[CLIENT].[…].list-manage.com/track/open.php?u=[…]&id=[...]&e=[ID] |
| Adestra | http://[CLIENT].msgfocus.com/t/[ID].png |
| MarkMonitor | http://cl.exct.net/open.aspx?[ID]&d=[…] |
| AppNexus | http://ib.adnxs.com/getuid?http://[…]/[ID]/[...] |
| Criteo | http://er.prod.verticalresponse.com/[…]/[ID]/pixel.gif |
| Litmus | https://[CLIENT].emltrk.com/[CLIENT]?d=[MAIL] |
| Optivo | https://tracking.srv2.de/op/[…]/[ID]-[ID]-[ID].gif |
| Bigfoot Interactive | http://pix.bfi0.com/t.gif?k=[…]&c=[…]&s=[ID] |
| Mailermailer | http://m1e.net/c?[ID] |
| VerticalResponse | http://cts.vresp.com/o.gif?[…]/[ID]/[…] |

## 6. Methodology

### 6.1. Experimental setup

To verify the effectiveness of the proposed features and the image-classification framework, we empirically test the accuracy of tracking image detection in a real-life environment. This prepares the development of a fully-functional tracking detection system, in which classification accuracy must be reliable over time and also perform well for senders not included in our sample. We approximate this performance by evaluating classifiers on three dimensions. We report performance on a typical test-set split from the training data, *out-of-sample*, and expand these results with an analysis of two additional test sets. The latter contain newsletters from the same companies sent after the training period, *out-of-time*, and from companies not in the training set, *out-of-universe*. Figure 6 summarizes the structure of the training and test setup including the size of the final data sets. The out-of-sample and out-of-universe test sets are drawn randomly from the data collected until October 31, 2015. The out-of-time test sets are emails received in 3-month-periods after the training period. The rest of this section describes the data sets in detail.

The training data consists of HTML emails received within a 5-month period between June 1 and October 31, 2015. It encompasses 215,565 images from 5,478 unique emails. For out-of-sample testing, we randomly select 548 (10%) of these emails and their images to evaluate models trained on the remaining data. The images in the test set are similar to the training data images in that they contain emails from the same time period as the training data and from senders contained in the training data. The sampling process is repeated ten times, respectively. Repeated testing ensures that results are reliable and not due to the random set of emails or companies selected for a single test set. This testing procedure is standard in the machine learning literature and comparable to the approach in Bender et al. (2016).

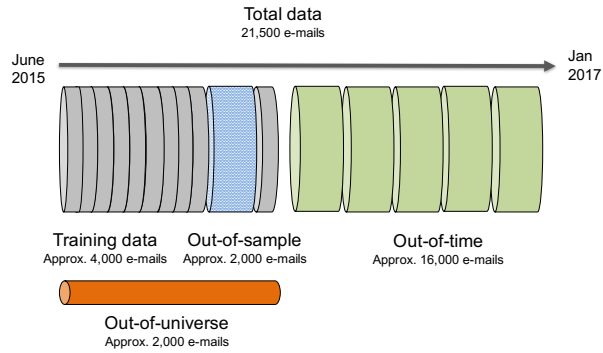

**Figure 6: Structure and size of the training and three test sets (random sampling of out-of-sample emails and out-of-universe companies repeated 10 times)**

To construct the out-of-universe test set, we randomly select 30 companies (identified by their sender domain) and assign all of their emails and images to the test set. Since no images from emails of these companies are used to train the model, the corresponding tracking infrastructure and reference structures are entirely unknown to the classifier. Testing on unknown tracking structures allows us to evaluate the performance of the final model on emails sent by different companies as an estimate of the performance of the classifiers on unknown senders in a real-world setting. In practice, the ability of the classifier to generalize to tracking infrastructures and senders beyond the 300 companies collected for this research is crucial. The sampling procedure is again repeated 10 times.

In order to capture a potential degradation of classification performance over time, we further define *out-of-time* test sets. The emails received after the training period, i.e. from November 1, 2015 until January 31, 2017, are divided into five sets each of which covers a three-month period. Since the content and structure of the emails and company infrastructure are expected to change over time, the results on the out-of-time sample provide an estimate of the performance of a static classifier after an extended period of time has passed. Since the difficulty of collecting ground-truth data inhibits frequent updating of the model or online learning, robust performance over time is an essential requirement to a reliable blocking approach.

As part of data preparation, we sample a subset of images from the training data for model estimation. The actual distribution of tracking images in the training data, which we use to build binary classifiers, may introduce two forms of bias. First, tracking images make up 8.1% of images in the training data leading to a skewed distribution between the target classes with potentially little variation within non-tracking images in a single email. Unbalanced target classes are known to cause an undesired focus of classification models on the majority class (Verbeke, Dejaeger, Martens, Hur, & Baesens, 2012). Second, the numbers of tracking images per email vary from 0 to 57 and thus differ substantially. Hence, tracking images of companies that include a large number of tracking images into their emails may be overrepresented in the data. This may introduce a sampling bias as a classifier may focus on frequent image structures sent by a small number of companies. To overcome these issues, we resample the training data through randomly selecting up to two tracking and content images, respectively, from every email in the training set. For emails containing less than two tracking or content images, respectively, all available images are selected. This approach excludes images from emails with a high number of tracking images and thus addresses the sampling bias. Our resampling also returns an approximately equal amount of tracking and content images for model training. To achieve this, it discards a sizeable fraction of content images, which suggest that our sampling approach can be considered a form of undersampling (Viaene & Dedene, 2004).

Regarding the application context of tracking prevention, it is important to take into account that the costs associated with different types of errors are uneven. Misclassifying an actual content image as tracking image, and thus filtering the image, may impede readability of the email and negatively influence user experience. On the other hand, misclassifying tracking images, and thus failing to block tracking, impedes user privacy. The proposed selective prevention system is intended to block specific images rather than all images in an email in order to inhibit the user experience as little as possible, while ensuring a maximum level of user privacy. If the cost ratio between false positives and false negatives can be specified, application specific costs can be included into model training, for example through increasing the ratio of target observations in the data via sampling or a reweighting of the model error (Viaene & Dedene, 2004). However, error costs appear an abstract construct in the case of tracking prevention. The misclassification costs depend on the personal risk assessment of an individual user and how she evaluates the relative severity of a privacy breach against the inconvenience associated with manual downloads of blocked images. Given these complications, we argue that a cost-sensitive model estimation is impractical in the focal application context and consider the cost imbalance through post-processing of model predictions described in Section 7.2.

## 6.2. Model Selection

We train and test several state-of-the-art machine learning algorithms to identify the binary classifier that is best suited to classify images as "tracking" or "non-tracking" based on the proposed features. Since prior work does not provide information on the performance of classifiers in this application, our selection of methods is based on classifier benchmarks in other domains (Lessmann, Baesens, Seow, & Thomas, 2015; Verbeke et al., 2012). All methods take numeric and categorical features as input to identify potentially non-linear patterns and produce a probability estimate of class identity given the feature values for an unknown observation. Each algorithm provides a number of tuning parameters, which describe, for example, the optimization behavior and complexity of the model. Table 4 provides a list of candidate models and parameters considered in the study. A comprehensive discussion of the classifiers is beyond the scope of the paper and available in, e.g., Hastie, Tibshirani, and Friedman (2002). We determine the best set of tuning parameters chosen using five-fold cross validation on the training set.

**Table 4: Classification methods and meta-parameter settings**

| Learning Algorithm |
|---|
| **Artificial Neural Network (Multilayer perceptron)** |
| Three-layered architecture of information processing-units referred to as neurons. Each neuron receives an input signal in the form of a weighted sum over the outputs of the preceding layer's neurons. This input is transformed by means of a logistic function to compute the neuron's output, which is passed to the next layer. The neurons of the first layer are simply the covariates of a classification task. The output layer consists of a single neuron, whose output can be interpreted as a class-membership probability. Building a neural-network model involves determining connection weights by minimizing a regularized loss-function over training data. |
| No. of neurons in hidden layer: 3, 5, …, 13 |
| Regularization parameter: $2^{[-4, -3.5, ..., 0]}$ |
| **Random Forest** |
| The ensemble consists of fully-grown CART classifiers derived from bootstrap samples of the training data. In contrast to standard CART classifiers that determine splitting rules over all covariates, a subset of covariates is randomly drawn whenever a node is branched, and the optimal split is determined only for these preselected variables. The additional randomization increases diversity among member classifiers. The ensemble prediction follows from average aggregation. |
| No. of member classifiers: 2000 |
| No. of covariates randomly selected for node splitting: 5, 8, 10, 12, 15, 20 |
| **Stochastic Gradient Boosting** |
| Modification of the AdaBoost algorithm, which incorporates bootstrap sampling and organizes the incremental ensemble construction in a way to optimize the gradient of some differential loss function with respect to the present ensemble composition. We employ tree-based models (CART) as member classifiers. |
| No. of member classifiers: 10, 25, 50, 100, 250, 500 |
| Learning rate: $10^{[-4, -3, ..., -1]}$ |
| Max. tree depth: 2, 4, 6, 8 |

Note that the table depicts only those meta-parameters for which we consider multiple settings. A classification method may offer additional meta-parameters.
We consider all possible combinations of meta-parameter settings for learners such as artificial neural networks that exhibit multiple meta-parameters.

The performance of the state-of-the-art classifiers is compared to three benchmarks. First, we employ a standard logistic regression model. This benchmark allows us to shed light on the trade-off between an interpretable linear model and more complex nonlinear classifiers, which are opaque but supposedly more accurate. Second, we consider the blacklist approach described above as a representative of a manually designed detection rule. This benchmark is to confirm the need for data-driven detection models. Third, we consider a static decision rule based on image size (area below 3 square pixels or not specified) and file format (categories *none*, *php* or *other*). Previous research (Bender et al., 2016) and our exploratory analysis finds these simple features to be highly predictive. It is thus interesting to check the detection performance of a corresponding classifier and whether it decreases over time. Clearly, the simple classifier, which we refer to as baseline model in the following, is vulnerable to even small adjustments by trackers. The manipulability of image size and displayed file format in particular disqualify this approach as a resilient, long-term solution. For the data employed here, examining the detection performance of the baseline model on the out-of-time data will shed some light on the degree to which an evolution of tracking practices has taken place over the observation period.

## 7. Empirical results

The quality of the detection model depends on its overall performance, generality, and resilience. We measure performance using statistical indicators of predictive accuracy, and generality as model performance under different experimental conditions. In the following, we analyze feature importance to determine the overall number and type of features on which the prediction of a detection model is based and relate these findings to the ease of feature manipulation. We then compare the performance of the models on the different test data sets in terms of their ability to detect tracking images while producing few false alarms. Both characteristics are important to maximize security and usability for the user, respectively.

### 7.1. Feature importance and resilience

An effective tracking blocker must be able to classify images that vary substantially from the images available for training. Since only a subset of potential senders can be sampled to collect ground-truth data, it is important that features generalize to unobserved senders. Furthermore, the detection engine should be resilient against efforts by trackers to modify their infrastructure to avoid detection. Before discussing the overall performance of the classifiers, we proceed with identifying the salient characteristics of tracking images and evaluate the strength of the proposed new features as determined by the models. Figure 7 shows the 15 top-performing features according to normalized feature importance averaged over all classifiers and presents their respective importance values for each classifier.

We use standard algorithm-specific methods to calculate the feature importance scores. For random forest and gradient boosted trees, the score corresponds to the relative improvement in the splitting criterion due to the split; calculated as the square root of the sum of squared relative gain of splits on the feature in a single tree and averaged over all trees in the model (Hastie, Tibshirani, & Friedman, 2009). For artificial neural network, the hidden-output connection weights of each hidden neuron are partitioned into components associated with each variable's input neuron using Garson's algorithm (Goh, 1995).

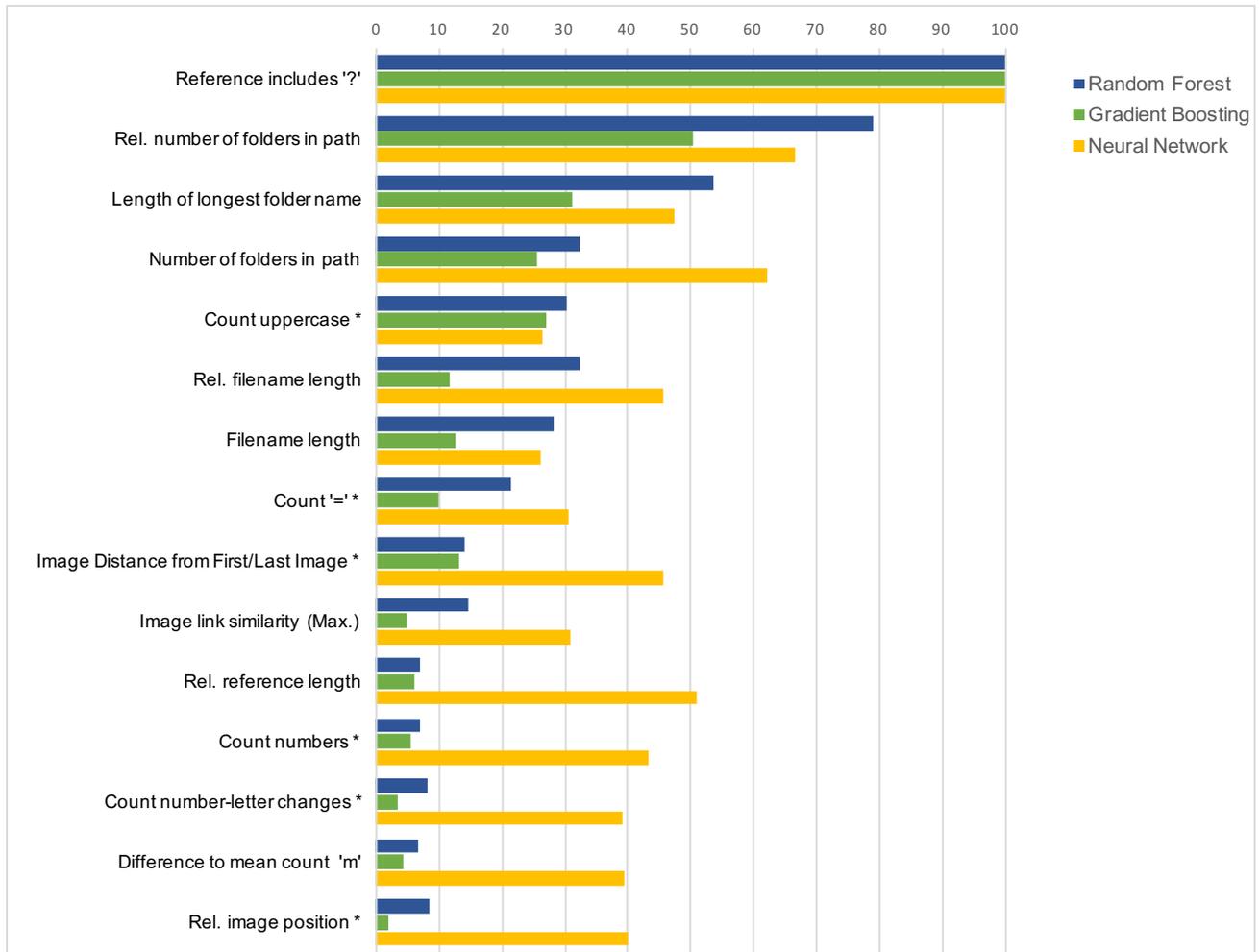

**Figure 7: The 15 most predictive variables selected according to average feature importance across all classifiers. Features marked with an asterisk have been introduced in Bender et al. (2016)**

Two main conclusions emerge from Figure 7. First, the ranking of variables is similar for all algorithms for the top feature after which there are substantial differences in the ranking between the tree-based models and the artificial neural network. This indicates that there are several highly predictive features within our selection of resilient features. All models rely heavily on the occurrence of a question mark, which indicates that parameters are passed on to a script and the folder structure of the image URL. Beyond the count of uppercase letters, the random forest model seems to consider a larger number of features than the gradient boosting model, e.g. the absolute and relative length of the filename. The neural network distributes importance more evenly and accounts for several defined patterns ignored by the tree-based models, e.g. the count of numbers or case-changes. It also places relatively large weight on email header characteristics, e.g. a match of image and sender domain. In practice, we would expect sparse models to generalize better to unknown data, but models considering a more diverse set of features to be more resilient to changes in tracking patterns.

Second, we observe that the novel features rank high in average importance over all classifiers and make up nine of the fifteen top features. These features are designed for deployability and resilience by capturing patterns that cannot be adjusted without a negative effect on the visibility of the tracking image or its tracking capability. On the side of technical restrictions, the occurrence of a question mark in the reference provides a convenient way to pass parameters to a tracking script and avoiding it would require costly changes to the data collection infrastructure. Even with an alternative solution, the existence of at least one unique identifier is necessary to map the image access to a specific email and email recipient. The existence of these IDs is captured by the (relative) length and number of folders as well as the length of the file name in the top features depicted in Figure 7. The large number of recipients requires a certain length and complexity of the ID, which consist of a random number and letter sequence. The randomness of these IDs is captured by the deviation in the number of times a letter is used in each reference to the average occurrence within the email.

On the side of organizationally costly adjustments, changes to the relative number of folders in the URL and all other relative measures regarding the reference structure require flexibility and coordination between different organizational units responsible for the management of content images and tracking images, respectively. For third party trackers, an additional issue is the implementation of changes to the existing server and folder infrastructure adjusted for each client, which requires the restructuring of existing systems. For example, the unification of reference folder length to hide tracking images, which commonly reside in very deep or very shallow folder trees, would require a standardized path structure set by the content management unit that still allows a convenient work environment and does not simultaneously increase the systematic deviation captured by the other relative features, e.g. length of folder name.

Given the high ranking of resilient features designed from technical restrictions and domain-knowledge, we expect the classifiers to be general and resilient with regard to unseen senders and changes over time, and against expected deliberate changes in the tracking infrastructure as outlined above. The following section evaluates the former two claims empirically.

### 7.2. Model performance

We evaluate classifier performance based on the area-under-the-ROC-curve (AUC), which captures a classifier's ability to discriminate between tracking and non-tracking images. We also use sensitivity and specificity statistics based on the optimal probability threshold (see Section 6.1) to evaluate the tracking detection accuracy of a classifier vis-à-vis its ability to not block content images.

The AUC allows us to summarize the performance of each classifier in a single metric aggregated over all potential thresholds and test the differences in performance statistically. The AUC for each classifier and test set, averaged over ten repetitions of random sampling, is given in Table 5. Note that the AUC is bounded between 0 and 1 (perfect discrimination), where a value of 0.5 corresponds to a random classifier. We also report the average ranks as the basis of a statistical analysis of model performance comparing the classifiers to the best performing classifier (Demšar, 2006). The last row of Table 5 depicts the test statistic and p-value of a Friedman test of the null-hypothesis that all classifier ranks are equal. Given that we can reject the null-hypothesis for all performance measures (p < .00), we proceed with pairwise comparisons of a classifier to the control classifier using the Rom procedure for p-value adjustment (García, Fernández, Luengo, & Herrera, 2010). Table 5 depicts the p-values corresponding to the pairwise comparisons in brackets. Italic face indicates that we can reject the null-hypothesis of a classifier performing equal to the best classifier (i.e., p < .05).

For all test sets, the random forest model performs best and thus serves as control model for statistical testing. We observe the benchmark models to perform significantly worse than the random forest classifier at the 5% level but are unable to establish a significant difference in performance between the random forest and the gradient boosting or neural network classifier. The results for the out-of-sample test set are comparable to previous studies and support the view that machine-learning classifiers are highly effective in identifying tracking elements (Bender et al., 2016). All non-linear classifiers achieve close to perfect performance and perform significantly better than the baseline model, which classifies images based on image size and file format. The blacklist model provides some discriminatory power. However, it performs significantly worse than the best alternative classifier.

Out-of-sample results represent the performance of a detection engine under ideal conditions. In practice, we cannot expect emails to originate from the same senders as in the training data. Additionally, the challenges in collecting labelled training data restrict model training to a relatively small number of different senders and impede regular re-training or updating of classifiers. We therefore evaluate the classifiers on out-of-universe test cases, which include only images from companies on which the model was not trained, and out-of-universe-and-time test cases, which include images from companies on which the model was not trained received after the end of the training period. As expected, we observe a decrease in AUC for all classifiers when applied to the more challenging test sets. This decrease is lowest for the random forest and gradient boosting models at 0.006 and 0.007 AUC points and highest for the logit model with a difference of 0.026 AUC points, suggesting that the tree-based ensemble models generalize well. Decreasing performance of the baseline model is surprising given its simple decision rule set and hints at a change in tracking practices in the out-of-universe/-and-time test sets. We attribute the fact that AUC increases for the blacklist model on more challenging test sets to sampling variance. The performance of the blacklist model is high whenever companies that use trackers from the blacklist are sampled for a random test set, and low otherwise.

Table 5: AUC and average rank classifier performance for each test set (10 sample average)

|  | Out-of-sample | | | Out-of-universe | | | Out-of-universe & -time | | |
|---|---|---|---|---|---|---|---|---|---|
|  | AUC | Rank | | AUC | Rank | | AUC | Rank | |
| **Blacklist** | 0.596 | 6.00 | *(0.00)* | 0.673 | 6.00 | *(0.00)* | 0.637 | 6.00 | *(0.00)* |
| **Baseline** | 0.969 | 5.00 | *(0.00)* | 0.953 | 5.00 | *(0.00)* | 0.938 | 4.80 | *(0.00)* |
| **Logit** | 0.998 | 4.00 | *(0.04)* | 0.982 | 4.00 | *(0.03)* | 0.972 | 3.95 | *(0.03)* |
| **Neural network** | 1.000 | 2.10 | (1.00) | 0.994 | 2.40 | (0.95) | 0.981 | 2.55 | (0.68) |
| **Random forest** | 1.000 | 1.95 | - | 0.997 | 1.80 | - | 0.994 | 1.75 | - |
| **Gradient boosting** | 1.000 | 1.95 | (1.00) | 0.996 | 1.80 | (1.00) | 0.993 | 1.95 | (0.81) |
| Friedman $\chi^2_5$ |  | 49.57 | *(0.00)* |  | 48.43 | *(0.00)* |  | 44.65 | *(0.00)* |

*Values in brackets give the adjusted p-value corresponding to a pairwise comparison of the row classifier to the best classifier (random forest). Italic face indicates significance at the five percent level. The last row shows the $\chi^2$ and p-values of a Friedman test to verify that at least two classifiers perform significantly different.*

The excellent discriminatory performance of classifiers, even on the out-of-universe-and-time test set, facilitates two conclusions. First, the features that we propose in Section 5 are sufficient to allow near perfect classification of tracking images within newsletter emails. This is important empirical validation that it is possible to identify tracking images without relying on image characteristics that are controlled by trackers. AUC values close to unity suggest that a tracking detection system built on resilient features can provide effective protection in the long run. Second, the proposed classification models generalize to newsletter emails received after the training period and from unknown companies. Generalizability is crucial due to the high number of potential senders and the difficulties in data collection outlined in Section 4, which impede frequent updating of the detection model.

Having established the predictive performance of the detection models, we examine the binary decision between loading and blocking an image in practice. This requires us to post-process the probabilistic predictions emerging from classification models. We obtain a crisp classification of images into tracking and non-tracking images through comparing probabilistic classifier predictions to a threshold. We then assess the accuracy of discrete class predictions in terms of the sensitivity and specificity of a classifier, i.e. the percentage of tracking and non-tracking images that are correctly classified, respectively. The definition of a threshold also offers an opportunity to account for uneven misclassification costs without having to specify actual cost values. We tune the probability threshold (for each classifier individually) on the training data set. Similar to applications in spam detection (Bergholz et al., 2008), medicine (Oztekin, Al-Ebbini, Sevkli, & Delen, 2017) and fraud detection (Viaene, Ayuso, Guillen, Van Gheel, & Dedene, 2007; Vlasselaer, Eliassi-Rad, Akoglu, Snoeck, & Baesens, 2016), the goal is to achieve a high detection rate with the lowest possible rate of false alarms. For the empirical evaluation, we define the probability threshold to be the value that maximizes the specificity of a classifier at a fixed sensitivity of at least 99.99% on the training data. We acknowledge the choice of 99.99% to be subjective. It is based on the believe that many users might have a strong preference for privacy and consider the misclassification of a tracking image to be the much more "costly" error compared to misclassifying a content image. Having fixed the sensitivity of each model on the training data, we compare the models on the most challenging out-of-universe-and-time scenario by comparing the sensitivity and specificity over ten random samples. In practice, sensitivity and specificity correspond to the ratio of detected tracking images and one minus the ratio of (erroneously) blocked non-tracking images, respectively. Results are presented in Table 6.

Table 6: Sensitivity and specificity of detection models across ten random test sets

|  | Out-of-universe & -time | | | | | | | | | | |
|---|---|---|---|---|---|---|---|---|---|---|---|
|  | 1 | 2 | 3 | 4 | 5 | 6 | 7 | 8 | 9 | 10 | Mean |
|  | Sensitivity | | | | | | | | | | |
| **Blacklist** | 0.03 | 0.19 | 0.35 | 0.23 | 0.47 | 0.28 | 0.15 | 0.29 | 0.41 | 0.51 | 29% |
| **Baseline** | 1.00 | 0.81 | 0.96 | 0.96 | 0.89 | 0.96 | 0.96 | 0.95 | 0.91 | 0.98 | 94% |
| **Logit** | 1.00 | 0.98 | 0.99 | 1.00 | 0.93 | 0.99 | 0.90 | 1.00 | 0.99 | 1.00 | 98% |
| **Neural Network** | 1.00 | 0.98 | 0.78 | 0.82 | 0.98 | 0.73 | 0.30 | 0.97 | 0.79 | 0.97 | 83% |
| **Random Forest** | 1.00 | 0.99 | 0.82 | 0.81 | 0.99 | 0.86 | 0.77 | 0.96 | 1.00 | 0.99 | 92% |
| **Gradient Boosting** | 1.00 | 1.00 | 0.86 | 0.85 | 0.84 | 0.88 | 0.80 | 0.96 | 1.00 | 0.99 | 92% |
|  | Specificity | | | | | | | | | | |
| **Blacklist** | 0.99 | 1.00 | 0.98 | 0.98 | 0.98 | 0.98 | 0.95 | 1.00 | 0.96 | 1.00 | 98% |
| **Baseline** | 0.94 | 0.94 | 0.96 | 0.92 | 0.95 | 0.90 | 0.95 | 0.93 | 0.93 | 0.98 | 94% |
| **Logit** | 0.54 | 0.93 | 0.78 | 0.59 | 0.80 | 0.74 | 0.89 | 0.76 | 0.56 | 0.59 | 72% |
| **Neural Network** | 0.96 | 0.98 | 1.00 | 0.99 | 0.98 | 0.98 | 1.00 | 0.99 | 0.94 | 0.99 | 98% |
| **Random Forest** | 0.90 | 0.98 | 1.00 | 0.99 | 0.97 | 0.99 | 1.00 | 0.99 | 0.95 | 0.98 | 98% |
| **Gradient Boosting** | 0.96 | 0.91 | 1.00 | 0.99 | 0.98 | 0.99 | 1.00 | 0.99 | 0.85 | 0.99 | 97% |

For all classifiers, we observe sensitivity to differ from our target value of 99.99%. Recall that this is the target value which we use to determine the classification threshold on the training data. Table 6 demonstrates that applying this threshold to unknown data decreases sensitivity (i.e., the accuracy of tracking image detection). Considering the tradeoff between high sensitivity and a low false alarm rate, Table 6 reveals that the random forest classifier has a higher tendency to sacrifice sensitivity for higher specificity compared to the logit model. We attribute the sharper decrease in sensitivity for random forest to the fact that random forest achieves almost perfect discrimination on the training data (Table 5), which leads to a higher, less strict classification threshold after optimization. Overall, the results suggest that the excellent discriminatory power observed in terms of AUC (Table 5) translates well to the actual decision problem under the proposed cutoff optimization scheme. When a user decides to allow loading external images for an email, the logistic regression or random forest classifiers robustly detect 98% and 92% of tracking images under the proposed system. At this level of performance, the detection models ensure a high level of user privacy under the most challenging conditions of an out-of-universe-and-time test. The negative effect on user experience is the false flagging of 28% and 2% of non-tracking images as tracking images, respectively. In the case of the random forest, we argue that the privacy gain outweighs the negative effect for users with even minimal preference for privacy. We judge the logistic regression under the proposed cutoff to be an alternative for users with a strong preference for privacy at the cost of a notable impact on user experience.

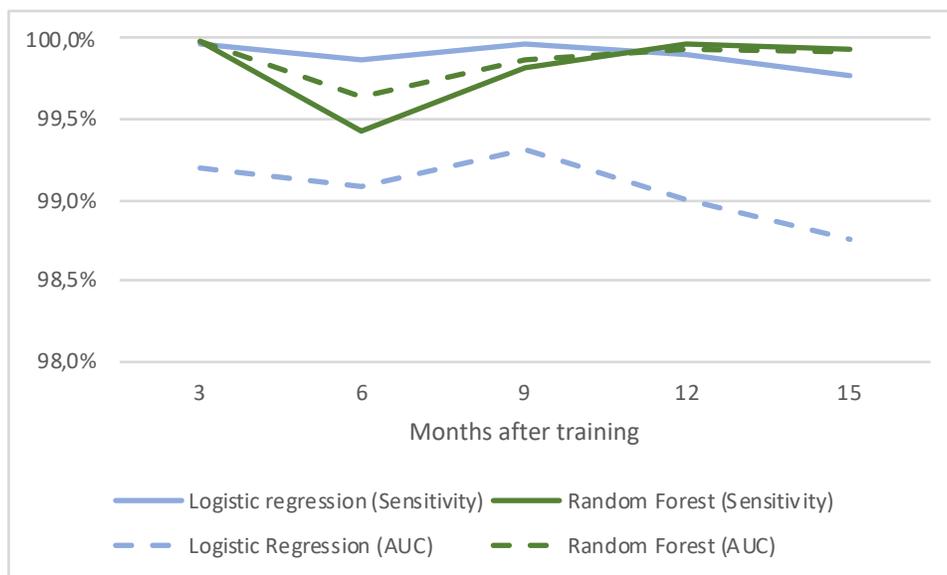

**Figure 8: Sensitivity after training period over five 3-month windows**

We further compare the dynamics of AUC and sensitivity of the selected models over time and determine a suitable interval for retraining each model. We conduct this analysis over five 3-month windows starting at the end date of the training data on out-of-sample test data to ensure sufficient sample size in each window. Figure 8 shows no trend in performance for the random forest and only marginal decrease in performance for logistic regression starting after nine months, with all changes within a 1%-interval of starting performance. We attribute the slump in performance for the 3-6-month window to particularities in the email schedule for the subscribed newsletter. Taken at face value, the results suggest that a detection engine preserves performance for a period of at least nine months after training, at which point the logistic regression could be re-trained on more recent data to avoid deteriorating performance. No long-term time effect is observed for the random forest within the observed period. These results are encouraging for practical applications where data collection and retraining are challenging. Furthermore, higher robustness of the random forest model supports the view that this model might be preferable to the logistic regression despite its lower sensitivity.

## 8. Conclusion

Email tracking can be used to gather identifiable and sensitive information on recipients without their consent or control, thus raising several security and privacy concerns. We describe the extent to which data can be collected that contains information on email reading behavior, system information, and location. In contrast to common web tracking, these data can be matched to an email address and, by extension, to the person behind the email account. Empirical analysis of over 30,000 emails from the 100 largest companies in Germany, Great Britain, and the United States, respectively, show that email tracking is widely applied. About 50% of all newsletter emails and close to 100% of emails in consumer-oriented industries include at least one tracking image. We identify the lack of a general, reliable and sufficient protection system against email tracking in previous literature and the software market, and propose a selective-prevention solution on the image level that is most suitable to balance privacy and usability.

We use the collected data to build a detection engine for the identification of tracking images based on machine learning. To achieve this, we outline a general methodology to infer resilient features from the technical characteristics of the tracking process. We follow this approach to design a comprehensive set of features that ensure applicability and resilience against tracker counter-strategies in a real-world setting. We test three state-of-the-art machine-learning classifiers and benchmark expected performance against heuristics proposed in previous research in a realistic application setting. In particular, we take into account long-term changes of tracking structures and classification of emails from unknown senders through repeated random sampling of three test data sets. We find a random forest classifier to provide the best overall classification performance at a detection rate of 92% and misclassification rate of non-tracking images at 2% for newsletters received from unknown senders after the training period.

Some caveats apply to the results gathered in this study, indicating directions for future research. The data used in this study contains commercial email newsletters, which exhibit important advantages for this research setting. Typical mail use involves additional mail categories, including private messages and the large category of spam and phishing emails. Further studies are required to check if our results generalize to tracking mechanisms in different types of email contexts. A fundamental threat to tracking image detection systems comes from the risk that actual content images could be employed for tracking. The proposed model could detect such images. However, their removal or blocking has a direct impact on the informational content of the email and thus conflicts with the interest of the user. Content-image tracking could be addressed using a server-side proxy solution. The server could cache all images with high tracking probability. Subsequent access to these (content and tracking) images from email recipients can then reference the server. Trackers would observe image downloads but only from the server so that the privacy of individual users is not compromised. The role of the proposed detection engine in a server-side solution would be to improve efficiency. Through selecting likely tracking images the server does not have to cache all images in all incoming emails. Furthermore, the problem of using content images for tracking is mitigated by the fact that tracking applications are often provided by specialized third-party services, for which the implementation of tracking mechanisms to content images within an email would require far more effort than attaching content-less tracking images. With such separation of content-management and tracking, an integrated solution will be costly for companies to realize. Further research on user behavior will prove useful to determine if users are willing to manually allow loading tracked content images.

With an extension of the data collection period, an analysis of changes to the features employed by the models and monitoring of model performance over time may provide insights into developments in tracking infrastructure and active countermeasures. Taking the long-term perspective, we have outlined the strategies that are available to trackers in order to actively hide tracking images from simple detection heuristics. Based on the available data and observation period, we come to the conclusion that the proposed detection system performs effectively and stable on the basis of the proposed resilient features. Consistently high tracking image detection rates on out-of-time and out-of-universe data suggest that the distribution of feature values or tracking practices has not changed during the observation period. Such change may however occur in the future. Therefore, future research to replicate our results and to perform a longitudinal analysis of feature distributions to collect evidence for a potential distributional shift seems highly relevant.

## 9. References

Agosti, M., & Di Nunzio, G. M. (2007). Gathering and mining information from web log files *Digital Libraries: Research and Development* (pp. 104-113): Springer.
Alsaid, A., & Martin, D. (2002). *Detecting web bugs with bugnosis: Privacy advocacy through education.* International Workshop on Privacy Enhancing Technologies.
Barret, B. (2015). A clever way to tell which of your emails are being tracked. *Wired*.
Bender, B., Fabian, B., Lessmann, S., & Haupt, J. (2016). *E-Mail Tracking: Status Quo and Novel Countermeasures.* Proceedings of the 37th International Conference on Information Systems (ICIS), Dublin, Ireland.
Bergholz, A., Chang, J. H., Paass, G., Reichartz, F., & Strobel, S. (2008). *Improved Phishing Detection using Model-Based Features.* CEAS.
Berners-Lee, T., Fielding, R., & Masinter, L. (1998). *Uniform Resource Identifiers (URI): Generic Syntax*: RFC Editor.
Blum, A., Wardman, B., Solorio, T., & Warner, G. (2010). *Lexical feature based phishing URL detection using online learning.* Proceedings of the 3rd ACM Workshop on Artificial Intelligence and Security.
Bonfrer, A., & Drèze, X. (2009). Real-time evaluation of email campaign performance. *Marketing Science, 28*(2), 251-263.
Bouguettaya, A., & Eltoweissy, M. (2003). Privacy on the Web: Facts, challenges, and solutions. *IEEE Security & Privacy, 99*(6), 40-49.
Cormack, G. V. (2008). Email spam filtering: A systematic review. *Foundations and Trends® in Information Retrieval, 1*(4), 335-455.
Demšar, J. (2006). Statistical comparisons of classifiers over multiple data sets. *Journal of Machine Learning Research, 7*, 1-30.

Dobias, J. (2010). *Privacy effects of web bugs amplified by web 2.0.* IFIP PrimeLife International Summer School on Privacy and Identity Management for Life.

Englehardt, S., Han, J., & Narayanan, A. (2018). I Never Signed Up For This! Privacy Implications of Email Tracking. *Proceedings on Privacy Enhancing Technologies, 1*, 109-126.

Evans, M., & Furnell, S. (2003). A model for monitoring and migrating Web resources. *Campus-Wide Information Systems, 20*(2), 67-74.

Evers, J. (2006). How HP Bugged E-Mail: Commercial online service was used to track email sent to a reporter in Hewlett-Packard's leak probe, investigator testifies. *CNET*. Retrieved from http://www.cnet.com/news/how-hp-bugged-email/

Fette, I., Sadeh, N., & Tomasic, A. (2007). *Learning to detect phishing emails.* Proceedings of the 16th International Conference on World Wide Web.

Financial Times. (2017). Equities, *Financial Times*. Retrieved from https://markets.ft.com/data/equities

Fonseca, F., Pinto, R., & Meira, W. (2005). *Increasing user's privacy control through flexible web bug detection.* Web Congress, 2005. LA-WEB 2005. Third Latin American.

García, S., Fernández, A., Luengo, J., & Herrera, F. (2010). Advanced nonparametric tests for multiple comparisons in the design of experiments in computational intelligence and data mining: Experimental analysis of power. *Information Sciences, 180*(10), 2044-2064.

Garera, S., Provos, N., Chew, M., & Rubin, A. D. (2007). *A framework for detection and measurement of phishing attacks.* Proceedings of the 2007 ACM workshop on Recurring malcode.

Goh, A. T. C. (1995). Back-propagation neural networks for modeling complex systems. *Artificial Intelligence in Engineering, 9*(3), 143-151. doi: https://doi.org/10.1016/0954-1810(94)00011-S

Harding, W. T., Reed, A. J., & Gray, R. L. (2001). Cookies and Web bugs: What they are and how they work together. *Information Systems Management, 18*(3), 17-24.

Hasouneh, A. B. I., & Alqeed, M. A. (2010). Measuring the effectiveness of email direct marketing in building customer relationship. *International Journal of Marketing Studies, 2*(1), 48-64.

Hastie, T., Tibshirani, R., & Friedman, J. (2002). *The Elements of Statistical Learning: Data Mining, Inference, and Prediction*. New York: Springer.

Hastie, T., Tibshirani, R., & Friedman, J. H. (2009). *The Elements of Statistical Learning* (2nd ed.). New York: Springer.

Hlatky, P. (2013). The Yesware Follow Up. Retrieved from http://www.yesware.com/blog/the-yesware-follow-up/

Hodgekiss, R. (2010). How Do I Create a Printer-Friendly Email Newsletter? Retrieved from https://www.campaignmonitor.com/blog/post/3232/how-do-i-create-a-printer-friendly-email-newsletter/

Javed, A. (2013). *POSTER: A footprint of third-party tracking on mobile web.* Proceedings of the 2013 ACM SIGSAC conference on Computer & communications security.

Jensen, C., Sarkar, C., Jensen, C., & Potts, C. (2007). *Tracking website data-collection and privacy practices with the iWatch web crawler.* Proceedings of the 3rd symposium on Usable privacy and security.

Kan, M.-Y., & Thi, H. O. N. (2005). *Fast webpage classification using URL features.* Proceedings of the 14th ACM International Conference on Information and Knowledge Management.

Kushmerick, N. (1999). *Learning to remove internet advertisements.* Proceedings of the third Annual Conference on Autonomous Agents.

Leon, P., Ur, B., Shay, R., Wang, Y., Balebako, R., & Cranor, L. (2012). *Why Johnny can't opt out: a usability evaluation of tools to limit online behavioral advertising.* Proceedings of the SIGCHI Conference on Human Factors in Computing Systems.

Lessmann, S., Baesens, B., Seow, H.-V., & Thomas, L. C. (2015). Benchmarking state-of-the-art classification algorithms for credit scoring: An update of research. *European Journal of Operational Research, 247*(1), 124-136. doi: 10.1016/j.ejor.2015.05.030

Li, M., Li, Z., Li, D., & Wang, B. (2011). Classification of images as advertisement images or non-advertisement images: Google Patents.

Ma, J., Saul, L. K., Savage, S., & Voelker, G. M. (2009a). *Beyond blacklists: learning to detect malicious web sites from suspicious URLs.* Proceedings of the 15th ACM SIGKDD International Conference on Knowledge Discovery and Data Mining.

Ma, J., Saul, L. K., Savage, S., & Voelker, G. M. (2009b). *Identifying suspicious URLs: an application of large-scale online learning.* Proceedings of the 26th Annual International Conference on Machine Learning.

Martin, D., Wu, H., & Alsaid, A. (2003). Hidden Surveillance by Web Sites: Web Bugs in Contemporary Use. *Communications of the ACM, 46*(12), 258-264.

Moscato, D. R., Altschuller, S., & Moscato, E. D. (2013). Privacy Policies on Global Banks' Websites: Does Culture Matter? *Communications of the IIMA, 13*(4).
Moscato, D. R., & Moscato, E. D. (2009). Information Security Awareness in E-commerce Activities of B-to-C Travel Industry Companies. *International Journal of the Academic Business World*, 39-46.
Murphy, K. (2014, December 24). Ways to Avoid Email Tracking, *The New York Times*. Retrieved from https://www.nytimes.com/2014/12/25/technology/personaltech/ways-to-avoid-email-tracking.html
Musciano, C., & Kennedy, B. (2006). *HTML & XHTML: The Definitive Guide*: " O'Reilly Media, Inc.".
Nikiforakis, N., Kapravelos, A., Joosen, W., Kruegel, C., Piessens, F., & Vigna, G. (2013). *Cookieless Monster: Exploring the Ecosystem of Web-Based Device Fingerprinting*. Paper presented at the Proceedings of the 2013 IEEE Symposium on Security and Privacy.
Oztekin, A., Al-Ebbini, L., Sevkli, Z., & Delen, D. (2017). A Decision Analytic Approach to Predicting Quality of Life for Lung Transplant Recipients: A Hybrid Genetic Algorithms-based Methodology. *European Journal of Operational Research*. doi: https://doi.org/10.1016/j.ejor.2017.09.034
Poese, I., Uhlig, S., Kaafar, M. A., Donnet, B., & Gueye, B. (2011). IP geolocation databases: Unreliable? *ACM SIGCOMM Computer Communication Review, 41*(2), 53-56.
Premkumar, G., & Roberts, M. (1999). Adoption of new information technologies in rural small businesses. *Omega, 27*(4), 467-484.
Ratcliff, J. W., & Metzener, D. E. (1988). Pattern Matching: The Gestalt Approach. *Dr. Dobb's Journal, 13*(7), 46.
Shih, L. K., & Karger, D. R. (2004). *Using urls and table layout for web classification tasks.* Proceedings of the 13th International Conference on World Wide Web.
Sophos. (2014). How emails can be used to track your location and how to stop it. *NakedSecurity*, from https://nakedsecurity.sophos.com/2014/02/27/how-emails-can-be-used-to-track-your-location-and-how-to-stop-it/
Technology Analysis Branch. (2013). What an IP Address Can Reveal About You: Office of the Privacy Commissioner of Canada.
The Direct Marketing Association. (2015). National client email report 2015.
Vaynblat, D., Makagon, K., & Tsemekhman, K. (2009). System and method for automatically delivering relevant internet content: Google Patents.
Verbeke, W., Dejaeger, K., Martens, D., Hur, J., & Baesens, B. (2012). New insights into churn prediction in the telecommunication sector: A profit driven data mining approach. *European Journal of Operational Research, 218*(1), 211-229. doi: 10.1016/j.ejor.2011.09.031
Viaene, S., Ayuso, M., Guillen, M., Van Gheel, D., & Dedene, G. (2007). Strategies for detecting fraudulent claims in the automobile insurance industry. *European Journal of Operational Research, 176*(1), 565-583. doi: 10.1016/j.ejor.2005.08.005
Viaene, S., & Dedene, G. (2004). Cost-sensitive learning and decision making revisited. *European Journal of Operational Research, 166*(1), 212-220.
Vlasselaer, V. V., Eliassi-Rad, T., Akoglu, L., Snoeck, M., & Baesens, B. (2016). GOTCHA! Network-based fraud detection for social security fraud. *Management Science*. doi: doi:10.1287/mnsc.2016.2489
Whittaker, C., Ryner, B., & Nazif, M. (2010). *Large-Scale Automatic Classification of Phishing Pages.* NDSS.
Yang, Y. (2010). Web user behavioral profiling for user identification. *Decision Support Systems, 49*(3), 261-271. doi: 10.1016/j.dss.2010.03.001